\documentclass{article}

\usepackage[T1]{fontenc}      
\usepackage[utf8]{inputenc} 
\usepackage[english]{babel}
\usepackage{amssymb,amsthm,amsfonts,amscd}
\usepackage{amsmath}
\usepackage{times}

\usepackage[authoryear,sort&compress]{natbib}

\usepackage{makeidx} 
\usepackage{graphicx}
\usepackage{algorithm}
\usepackage{algorithmic}
\usepackage{natbib}
\graphicspath{{.},{img/}}
\usepackage{url}

\usepackage{breakurl}
\usepackage[breaklinks]{hyperref}

\makeindex

\usepackage{vmargin}
\title{On the Morality of Artificial Intelligence}

\author{
  Alexandra Luccioni and Yoshua Bengio\thanks{Also CIFAR Senior Fellow} \\
  Universit\'{e} de Montr\'{e}al, Mila}
\date{December 2019}

\begin{document}

\maketitle

\begin{abstract}
    Much of the existing research on the social and ethical impact of Artificial Intelligence has been focused on defining ethical principles and guidelines surrounding Machine Learning (ML) and other Artificial Intelligence (AI) algorithms~\citep{jobin2019, ieee2017}. While this is extremely useful for helping define the appropriate social norms of AI, we believe that it is equally important to discuss both the potential and risks of ML and to inspire the community to use ML for beneficial objectives. In the present article, which is specifically aimed at ML practitioners, we thus focus more on the latter, carrying out an overview of existing high-level ethical frameworks and guidelines, but above all proposing both conceptual and practical principles and guidelines for ML research and deployment, insisting on concrete actions that can be taken by practitioners to pursue a more ethical and moral practice of ML aimed at using AI for social good.
\end{abstract}


\section{Artificial Intelligence Leaves the Research Lab}
Progress in ML in the last decade has been extraordinary and has rekindled the notion that AI systems could eventually reach human levels of performance, which was abandoned for several decades. Even if we are still currently far from this achievement, technological progress in ML has passed a threshold which enables it to have a huge economic impact, estimated to be close to 16 trillion US dollars by 2030~\citep{economicai2019}. This contrasts with the first few decades of ML progress, when researchers had the luxury of focusing purely on the fundamental aspects of their work, not worrying too much about its potential societal impacts -- an object recognition algorithm could be tested on a common dataset like MNIST~\citep{lecun1998} or ImageNet~\citep{deng2009}, and an objective performance metric would be obtained in order to measure progress, without having to think about the messiness and complexity of deployment and social impact. Something crucial has changed in recent years, as algorithms initially developed in the lab are increasingly being improved and deployed in society, in real-world applications such as healthcare, transportation and industrial production with real-life consequences, and we are likely seeing just the tip of the iceberg in terms of social impact. 
Along the way, this deployment in society has forced the realization that these algorithms have social impacts which could be positive or negative. For example, we have realized that biases hidden in data and algorithms could lead to more discrimination, in the simplest case simply because of the data imbalance: facial recognition algorithms have been found to underperform on gender and racial minorities~\citep{buolamwini2018}. Furthermore, above and beyond hidden biases, given the high impact potential of ML research, the question stands of whether practitioners are acting with the best interests of humanity and society in mind when developing their tools and applications. 

As ML researchers and engineers, we believe that we have a shared responsibility to consider both ethics and moral values when we choose what we work on, for what organization, and whether the products we contribute to directly or indirectly will be beneficial to humanity or more likely to end up hurting more than helping. Unfortunately, very few of us have been trained to think about these questions. Instead, most of us have focused from a very young age on mathematics and computer science and not so much on philosophy and other humanities. A good step towards learning about these issues is to consult the documents proposing ethical guidelines for AI, which we will cover in Section~\ref{existing}. Furthermore, in order to offer a guiding direction for such debates and soul-searching within the scope of ML, we propose the following self-directed questions:
\begin{enumerate}
    \item How is the technology that I am working on going to be used?
    \item Who will benefit or suffer from it?
    \item How much and what social impact will it have?
    \item How does my job fit with my values? 
\end{enumerate}

We are conscious that the questions listed above are subjective and the answers will depend highly on the values and ethics of the individual answering them. Nonetheless, we hope that work on some applications, such as the design and deployment of lethal autonomous weapons and automatic surveillance, will clearly be seen to contradict fundamental rights and dignity, as defined in, among other places, the UN Declaration of Human Rights~\citeyearpar{un1948}. Other applications of ML, such as those increasing the efficiency of advertising or beating the stock market, are less clear-cut in their moral value, and merit informed debate and discussion within the scientific community and society at large. As some of us become more conscious of the potential or definite social impact of ML, we have the opportunity, if not the duty, to make our voices heard. A good example of this is \href{https://futureoflife.org/open-letter-autonomous-weapons/}{a recent letter} signed by numerous scientists calling for an international treaty banning lethal autonomous weapon systems, e.g., killer drones which can decide to shoot at a person without human involvement, which would make it possible to take the broad social, moral and psychological context into account and potentially decide to abort the mission (for instance, when the target is in a school or at a family dinner surrounded by women and children).

Finally, while the legal frameworks to oversee and limit research and development violating these principles are often and unfortunately updated in a reactive rather than a proactive manner, we believe that we should not wait until all of the dots between ML and ethics are formally connected by legislation and regulation. We believe that we have a responsibility to educate ourselves, to think ahead about potential consequences, to use our internal moral compasses and to consciously choose the direction of the research or engineering that we practice. This is important because we believe that we are faced with a \emph{wisdom race}: as technology becomes more powerful, its impact can be proportionally greater, either positively or negatively.  

To curb the negative impact, we need to become wiser individually (as reflected in our personal decisions) and collectively (through social norms, laws and regulations). Unfortunately, technological progress in AI has accelerated faster than the current rate of progress of personal and social wisdom, ultimately making it possible for unwise humans or organizations, even those with good intentions and acting legally, to have large-scale, major destructive effects. This is comparable to a world in which nuclear bombs (i.e. very powerful technology) were accessible and usable by children (i.e., persons with insufficient maturity and wisdom), which could easily result in global nuclear war. This highlights the importance of the discussions still to be had by large numbers of ML practitioners about ethics and social impact, as well as the safeguards that need to be put in place to protect especially the most vulnerable members of our society. We will discuss some of the most advanced efforts to introduce these safeguards in the next section, followed by some examples of socially beneficial applications of ML.

\section{Ethics and AI - Existing Initiatives}\label{existing}

In recent years, there have been numerous initiatives which have taken one of two major approaches to fostering the ethical practice of AI: (1) Proposing principles guiding the socially responsible development of AI or (2) Raising concerns about the social impact of AI. We will describe both approaches in the current section, as well as giving examples of notable initiatives and projects which have adopted either of the approaches.\footnote{For a more complete overview of different global ML ethics initiatives, see a recent review in~\cite{jobin2019}}

\subsection{Defining Principles for Practicing AI Responsibly}

The topic of ethical research and practice in technology has been gaining momentum in different corners of the computing community in recent years, and the various initiatives that have been proposed are indicative of the interest and the concern that many members share. For instance, in the United States, the Association for Computing Machinery (ACM) has proposed a Code of Ethics and Professional Conduct, to be followed by all members of the association and to guide them in their usage of computer science~\citep{gotterbarn2018}. A similar initiative was undertaken by the Royal Statistical Society (RSS) in the United Kingdom, which has created a practical guide for practitioners regarding the ethical use of mathematics~\citep{rss2019}. In the present section, we will address the two most relevant and extensive initiatives to establish ethical guidelines for AI research and practice: the Montreal Declaration for Responsible Development of AI and the IEEE report for Ethically Aligned Design.

\subsubsection{The Montreal Declaration for a Responsible Development of Artificial Intelligence}

One of the most notable approaches to establishing guidelines for AI deployment is the Montreal Declaration for a Responsible Development of Artificial Intelligence, developed in 2017 and revised in 2018 based on public feedback\footnote{ \url{https://www.montrealdeclaration-responsibleai.com/}}. It was elaborated under the premise that given the assumption that since AI will eventually affect all sectors of society, it requires principles to guide its development to ensure its adherence to human values and social progress. The resulting Declaration has ten principles, ranging from protection of privacy to equal representation, with some principles touching responsibility and ethics directly; for instance, the principle of Prudence stipulates that \emph{“Every person involved in AI development must exercise caution by anticipating, as far as possible, the adverse consequences of AIS [Artificial Intelligent Systems] use and by taking the appropriate measures to avoid them.''} These principles were defined after extensive debates and dialogue between both specialists and non-specialists from different domains and parts of the world to ensure representability and cohesion. The overall aim of the declaration was to spark public debate and to encourage a progressive and inclusive orientation to the development of AI.

However, the Montreal declaration goes further than theoretical ethical principles, proposing recommendations to accomplish an ethical digital transition that includes all of the different levels of society, from researchers to policy-makers. For instance, it includes a proposition for auditing and validating the use of AIS using concrete frameworks and certifications in order to prevent biases and discrimination. Specific steps were also proposed for ensuring the protection of democracy and reducing the environmental footprint of AI, all within the framework of a democratic and citizen-led process. This is important given that the effects of AI will permeate all levels of society, from the programmers and engineers who write the code, to the leaders who will legislate it, and the businesses who will make products with it that will be used by all. The process of creation of the Montreal declaration was consequently the keystone to building a way of including all of these different stakeholders in the elaboration of an ethical AI, and paves the way for subsequent work on the topic.

\subsubsection{IEEE Ethically Aligned Design}

A more recent effort, initiated by the IEEE Global Initiative on Ethics of Autonomous and Intelligent Systems, carried out an in-depth study on the issue of the ethics surrounding the design of AI systems~\citep{ieee2017}. In particular, aspects that are relevant to the topics covered in the present paper include: the usage of A/IS [autonomous and intelligent systems] in service to sustainable development for all, and more specifically for the attainment of the United Nations Sustainable Development Goals (SDGs)~\citep{sdg2017}. The authors of the study specifically underline the potential of AI to contribute to resolving some of the world’s most urgent problems, such as climate change and poverty, given the necessary will and orientation towards these problems. Furthermore, they highlight the fact that despite their great potential, current AI deployment and development is currently not aligned with these goals and impacts~\cite[p.~144]{ieee2017}, which is unsettling given the myriad of ML project and initiatives worldwide.

The IEEE report also lays down principles to guide \emph{“the ethical and values-based design, development, and implementation of autonomous and intelligent systems''}, many of which are similar to those defined by the Montreal Declaration: respect of human rights, data agency, transparency, accountability, etc. They go further in proposing that \emph{“A/IS creators shall adopt increased human well-being as a primary success criterion for development”} instead of focusing on isolated metrics such as accuracy, and, from a deployment perspective, offering alternative metrics to quantify meaningful progress, for instance by evaluating social, economic and environmental factors instead of profit and other common success metrics. The report also includes propositions for policymakers, legislators and other stakeholders from the extended AI community and, as such, represents the most extensive effort of establishing ethical boundaries and guidelines for AI research to date.

In a recent survey of the various global ethics guidelines proposed around AI, the authors observed that despite a conceptual overlap between the many existing guidelines, including the two mentioned above, there are major differences regarding how the principles are interpreted~\citep{jobin2019}. This underlines the complexity and nuance of applying theoretical, philosophical principles in practice, and raises questions such as: what aspects of the AI research and deployment pipeline do ethics principles affect? how would it be possible to resolve conflicts between, for instance, fairness and sustainability (i.e. training an algorithm longer and with more data - thus potentially leading to more greenhouse gas emissions - to ensure that it is not discriminatory and covers all demographic groups equally well)? And, above all, how is it possible to translate ethical principles into a programming language? In any case, the bridge between theory and practice has yet to be built and there are different ways in which that can happen. This underlines the necessity of involving actors from different levels of the AI ecosystem (and neighboring ones) in order to ensure that experts in policy-making work in tandem with experts in coding and engineering to create tools and frameworks that are coherent and usable by all.

\subsection{Identifying Ethical Concerns of AI Applications}

There are several types of ethical  concerns regarding AI applications and, in this paper, we will focus more concretely on bias leading to potential discrimination. While it is true that on the one hand, AI-infused technology such as computer vision can enhance public security, for instance by identifying crime in real-time based on CCTV cameras, but the trade-off is that can also be abused to track individuals and to establish a surveillance state where privacy is greatly threatened by those who control the technology. On the military side, similar technology can be used to design autonomous drones which use computer vision to identity their target, representing a grave threat to global security and democracy due to the lack of human oversight. In addition to the security risk, such weapons would be moral and legal hazard: AI technology is not yet capable to comprehend and represent the social and psychological context in which such a targeted attack could take place in a manner that is coherent with international laws regarding war as well as with human morality. 

Unfortunately, the most common argument brought in favour of developing lethal autonomous weapons is that they are needed as a precautionary measure (i.e. since other countries are undeniably working on them, each country needs to do the same). In reality, the weapons needed to defend against killer drones would be very different from the drones themselves, and do not need to be lethal autonomous weapons since they would be designed to destroy weapons rather than to target people, similar to the Iron Dome used by Israel. Another common argument is that an international treaty would be useless since some countries will refuse to sign it. But we have seen in the past that even when major powers do not sign a treaty (such as the one on anti-personnel mines, signed by 133 countries, excluding the US, in 1997), the treaty can still be used to create a moral stigma, as well as a decline in demand; in the case of anti-personnel mines, the result has been that U.S. companies have stopped building them, even though their government never signed the treaty. Another flawed argument is that regulating lethal autonomous weapons could threaten the innovation in AI, whereas in fact AI has been developed very successfully in a civilian setting (mostly in academia and major technology companies) and its continued development does not require neither data nor engineering which would come from AI military development.

Another potential threat to democracy stemming from AI could come not simply from the increased ability to monitor and to target individuals, but also from the more subtle power to influence them, e.g. via AI-driven advertising, automated online trolls and other psychological manipulations via the internet and social media. The recent use of AI to influence political campaigns such as the 2016 US election or Brexit is just the beginning of what can be done when machines learn how to “press our buttons'' in a personalized way. This is due to the fact that micro-targeting makes it possible for ads to be truly bespoke depending on your political views, network of friends and personal history. While we may not mind being influenced when it comes to choosing a brand of soft drinks, when the profit or power motives of a corporation or political organization go against our individual and collective interests, it becomes important to establish social norms, laws and regulations to protect us from such psychological manipulation. But where should the line be drawn between, for example, manipulation and education? These are difficult questions but there are clues which can be used (like whether the organization that stands to profit is paying for the advertisement or social network influence), so human judgement remains key for judging the ethical aspect, e.g.  in balancing different values (like autonomy vs well-being, when considering an ad campaign against cigarettes, for example). In the case of advertising, what is interesting is that in addition to the moral hazard associated with psychological manipulation, it is not even clear that advertising is beneficial to society from a purely economic perspective, as it tends to favour established brands and thus slow down innovation.

Closely related to the political misuse and manipulation with AI is also increasing concern about AI-generated false images, videos and news. Thanks to rapid progress in generative neural networks such as the GANs~\citep{goodfellow2014}, it is becoming possible to synthesize images and sounds in a controlled way, e.g., using “deep fakes'' for making a video of a president declaring war, or with the face of a celebrity seamlessly integrated on the body and behavior of a pornography actor. Other commonly discussed concerns of AI deployment include the effect on the job market~\citep{wef2018}, which means that governments and communities must prepare, e.g. by adapting the education system and the social safety net, which can take decades, as well as the potential concentration of power which it may lead to in specific individuals, corporations and countries, and the bias and discrimination it may contribute to increase, as we discuss next. 

\subsubsection{Identifying and Mitigating Bias}

In recent years, we have been confronted numerous times with the fact that biased algorithmic systems can perpetuate injustice and discrimination, whether we are aware of it or not. There are many different ways that this kind of bias can creep into algorithms: it can be from the data itself, or the implicit bias that the creator programmed into the system, and even the way the problem is framed\footnote{For a more hands-on presentation of bias and fairness in AI, we suggest \href{https://developers.google.com/machine-learning/crash-course/fairness/video-lecture}{Google’s Online course designed specifically for ML practitioners}}. Therefore, in order to ensure that the models that we develop and the systems that they are later used in are as fair and ethical as possible, there are steps to take to identify bias and to reduce it as much as possible.

\paragraph{Numerical Bias}\mbox{} \\

A major challenge in designing ML systems is understanding how they work during training and deployment, and what factors and features they use to make decisions. However, diagnosing the presence of bias in these systems is not a straightforward task, since it is not always obvious during a model’s construction what the downstream impacts of design choices may be; therefore, upstream efforts are needed to reduce this risk as much as possible. To this end, there have been several proposals to help practitioners identify and mitigate bias in ML models, some of which we will describe in the current section. 

More concretely, exploring, analyzing and visualizing the data used for training a model is a key part of the ML process. But it is not straightforward to identify bias simply by looking at the data; often, more in-depth probing is needed to figure out what features and implicit information is present and, once a model is developed, how this will influence the model’s behavior. For instance, it was recently found that the COMPAS system, a criminal risk assessment tool developed widely used in the United States, is often biased with respect to race~\citep{angwin2016}. Whereas the bias in the COMPAS system was identified after its deployment, once the data was made public, this bias is an aspect of the model that should have been identified much earlier, during development and certainly before deployment. Similarly, off-the-shelf facial recognition technology used by police forces has been shown to perform much worse on racial and gender minorities, with a difference of up to 34.4\% in error rate between lighter-skinned males and darker-skinned females, mostly due to the lack of reliable training data~\citep{buolamwini2018}.

To address these types of issues, several approaches exist: for instance, researchers have recently released a tool called ‘What-If’, an open-source application that lets practitioners not only visualize their data, but also test the performance of their ML model in hypothetical situations, for instance modifying some characteristics of data points and analyzing subsequent model behavior, by measuring fairness metrics such as Equal Opportunity and Demographic Parity~\citep{wexler2019}. Other approaches address bias by changing the training procedure or the structure of ML models themselves, for instance by transforming the raw data in a space in which discriminatory information cannot be found~\citep{zemel2013} or using a variational autoencoder to learn the latent structure from the dataset and using this structure to re-weight the importance of specific data points during model training~\citep{ribeiro2016}. Whatever the approach chosen, using these kinds of tools during ML model development and deployment can change the life of individual people, who could go from unfairly spending decades in prison to having the chance of a better life -- an immensely important difference when multiplied by the thousands of people whose lives can be affected by the deployment of these tools. This multiplication of bias is especially important to consider since ML is being used more and more, and therefore even edge cases and small minorities can be amplified in real-world applications.

\paragraph{Textual Bias}\mbox{} \\

Bias is not always in numbers, it can also manifest itself in the words that we use to describe the world around us. For instance, in 2018, Reuters reported that Amazon was forced to decommission an ML-powered recruiting engine when it was discovered that it penalized any mention of female-related vocabulary, including applicants who attended all-women colleges~\citep{dastin2018}. This is not surprising given the gender disparity that exists in the technology sector and since the data used to develop this tool was comprised of resumes submitted (and accepted) to Amazon over a 10-year period. It is nonetheless disturbing in terms of algorithmic fairness, especially if algorithms such as this one make filtering or hiring decisions that can ultimately affect an entire gender’s lives and careers. This can potentially create a negative feedback loop, as such a system would reduce the number of female workers and thus the number of positive role models for girls interested in technology. A similar type of gender bias was also found in pretrained word embedding models, which were found to exhibit gender stereotypes in terms of higher cosine similarity between, for instance, ‘woman’ and ‘homemaker’ or ‘receptionist’ as opposed to ‘woman’ and ‘doctor’ or ‘lawyer’, notably due to these biases existing in the corpus that they were trained on, which consisted of mainstream news articles~\citep{bolukbasi2016}.

In order to reduce and eventually remove gender bias in written text, researchers have proposed approaches such as identifying the gender subspace of vectors and adjusting the dimensions in a way that either neutralizes or entirely removes gender bias~\citep{bolukbasi2016}. Others have defined a formal gender bias taxonomy in order to capture gender bias and to train ML models to later identify this bias in texts~\citep{hitti2019}. Debiasing the computational representation of language, notably word embedding models, is especially important because of the extent of their usage; pretrained embedding models trained on corpora such as Google News and the Common Crawl are used in a variety of applications and systems, and can therefore continue perpetuating gender bias in downstream usages in Natural Language Processing (NLP) applications such as dialogue systems. This is a challenge given the complex and sub-symbolic nature of modern NLP, which makes it difficult to analyze specific features and aspects of data and identify latent connections and bias between words and concepts. Therefore, more work is needed to explore and analyse these issues, which constitutes an interesting research direction in itself, and one that is important to pursue and to integrate into mainstream ML research.

Despite the research initiatives described above to carve appropriate social norms about AI, there remains a noticeable gap between the recommendations they make and ways to ensure that these are respected. Legislation of AI is still catching up to the progress made in research and practice, and there have not yet been any country-level laws governing AI research specifically. However, there have been, on the one hand, more high-level legislative frameworks such as the \href{https://gdpr-info.eu/}{European Union (EU) General Data Protection Regulation (GDPR)}, which aims to ensure data privacy and protection and, on the other hand, more local initiatives such as \href{https://sfgov.legistar.com/View.ashx?M=F&ID=7206781&GUID=38D37061-4D87-4A94-9AB3-CB113656159A}{San Francisco’s Facial Recognition Software Ban}. Nonetheless, more complete legal frameworks are needed to control nefarious use of AI and to ensure that the principles defined in theory are applied and enforced in practice.

\section{AI for Good Initiatives}

Whereas the profit motive is the main driver behind much of the commercial deployment of AI today, there are nonetheless many projects going on in academia, government organizations, civil society and industry labs motivated by more noble objectives, often called AI for Social Good (AISG) projects. In addition to the specific projects being undertaken in areas such as healthcare, education or the environment, it is interesting to highlight higher-level efforts which aim to foster and facilitate these projects. For example, the \href{ https://ai-commons.org/}{ AI Commons projec} aims to construct a hub where different kinds of actors can connect and collaborate on AISG projects, e.g., ML graduate students or engineers, problem owners in NGOs or local governments,  philanthropy organizations, or startups which could deploy the ML solutions. Their interaction is to be facilitated by online tools and datasets as well as a standardized description of the status, progress and expected impact of each project. We hope that initiatives like this will help solidify and amplify the impact of AISG; in the meantime, there are also many profoundly positive uses of AI that are emerging and we would like to highlight and applaud such efforts in the present section. 

\subsection{AI in Healthcare}
  
Achieving universal health coverage is one of the 17 UN Sustainable Development Goals~\citep{sdg2017} and although major progress has been made in numerous domains, such as maternal health as well as HIV/AIDS reduction, there are still many problems that are far from being solved. While ML is not a cure-all, there are many challenges that it can help with such as personalized medicine, diagnosis of medical imagery, and improved drug discovery~\citep{ghassemi2018}.  ML in the health sector is in fact a thriving domain of research, with its own workshops at major ML conferences and research published in major medical journals read by practitioners worldwide. In the last five years alone, groundbreaking work has been done in improving the diagnosis of diabetic retinopathy from a single visit~\citep{arcadu2019}, detecting breast cancer in lymph nodes~\citep{golden2017}  and large-scale discovery of diseases based on health records~\citep{pivovarov2015}. There is also an increasing number of startups and companies working in the space, either by commercializing research done in academia or by developing products specifically catered to the medical sector, with the most advanced applications harnessing the power of deep learning for analyzing and classifying medical imagery.

Despite the many exciting advances that are being made, there are many hurdles in ML research in healthcare, starting from data privacy and control (who owns the data? Can patients share their own data, or should the process be centralized? How to find the right balance between privacy and the lives which will be saved by applying ML on the aggregated health records from many different sources?), to the manner in which medical data should be processed (Should it contain information such as race and postal/zip code, which can impact diagnoses, be included in electronic heath records, or does that open the door to discrimination and bias?) and how should such systems be deployed (human-in-the-loop or fully automated?)\footnote{For a more extensive overview of the opportunities and challenges of using ML in healthcare, see~\cite{ghassemi2018}}. There are also often questions of responsibility and interpretability that arise, given the high stakes of deploying ML systems in situations of life and death. In order to make meaningful progress in this sector, it is therefore important to continue existing research on fair and ethical usage of ML in healthcare~\citep{wiens2019} and to ensure that Hippocratic principles are a solid part of the research and development process, as well as working with stakeholders of the domain (e.g. radiologists, clinicians, patient organizations and hospital administrators) to propose solutions to the hurdles proposed above. 

\subsection{AI for Education}

The promise of using adaptive intelligent systems and agents for education has been around since the 1960s~\citep{suppes1969}, but access to personalized digital education tools has yet to become a reality in most countries, especially in the developing world, where it could have the most impact to democratize education and knowledge~\citep{nye2015}. In recent years, given the increasing global shortage of qualified teachers along with the increasing number of students, the issue of access to education has become a global one, a fact highlighted by its presence among the UN SDGs. And yet, the usage of ML in the education sector has been limited to specific, narrow applications such as predicting the probability of learner attrition~\citep{chaplot2015} or improving learner evaluation~\citep{abbott2006}. There are many reasons for this, starting from the difficulty in representing learning content in a domain-agnostic way to facilitate scalability, to overcoming cultural and linguistic barriers to deploying tutors worldwide, but also more fundamental issues such as the lack of large-scale educational datasets and the inherent technological constraints in developing countries. 

Despite these hurdles, there are many new and longstanding efforts to create intelligent tutors, be it using symbolic AI approaches such as ontologies and knowledge modeling~\citep{nkambou2010}, educational data mining~\citep{dutt2017} or, more recently, ML-driven approaches~\citep{conati2018}. However, there are very high stakes in the field, since technological interventions have the potential to make considerable, long-term impact on human livelihoods, for example lifting people out of poverty by endowing them with linguistic and numerical literacy, but these can be hindered by bias and technological constraints. We therefore agree with recent proposals to improve and support human learning at scale and believe that ML has a key role to play in this endeavour. This can be done, for instance, by partnering up with existing education initiatives and organizations in order to learn what their specific needs are and how ML can be used to meet them, or else by collaborating with Massive Open Online Course (MOOC) creators in order to gather data and make it available to the ML community, and finally by sharing learning materials and activities used in local education initiatives (e.g. university courses in Machine Learning) so that they profit learners in places where access to high-quality technical education is limited.

\subsection{AI for the Environment}

Climate change is, without a doubt, one of the biggest challenges that humanity has faced, and we are at an important point in history when we are both aware of the issue and still have the possibility to change its course. Climate change has been described as a ‘wicked’ problem, due to features such as the difficulty in defining the problem itself and in developing and deploying solutions to it, the lack of central authority that can solve it, the incentives for individual countries or companies to not do their share, and the cognitive biases that discount the future impacts of our actions~\citep{head2008,levin2012}. Furthermore, while we do not know of any single technological silver bullet as solution to climate change, there are nonetheless numerous technical challenges for which ML can be helpful, and which can be combined to make a significant impact on the overall issue. These challenges and the ongoing ML approaches to tackle them were presented in a recent survey paper~\citep{rolnick2019}. We will not go into all of these at length, but we will focus on a few examples that are particularly salient and that we hope will give an idea of both the relevance of deploying ML in environmental applications and the opportunities that this can generate.

\paragraph{Energy and Transportation}\mbox{} \\

Together, electricity and transportation systems are estimated to produce close to half of anthropogenic greenhouse gas (GHG) emissions~\citep{allen2019} and both sectors have their own unique challenges for decarbonization. For instance, one of the major obstacles to building and using renewable energy sources such as solar and wind is the variability of their output, which is inherently problematic since the power generated by an energy grid must equal the power used by its consumers at any given moment. Currently, this means that despite the existence of solar panels and wind turbines, these must be complemented by controllable but highly polluting energy sources such as coal and natural gas plants. ML methods that are appropriate for time-series predictions, such as Recurrent Neural Networks are particularly suited for these types of tasks~\citep{voyant2017} and can dramatically lower the barrier to entry for renewable energy globally. Furthermore, even in cases where controllable energy sources are used, demand on the energy grid will still fluctuate based on usage; in this case, ML techniques such as Reinforcement Learning and Dynamic Scheduling can be used to balance the grid in real time~\citep{vazquez2019}.

In transportation, reducing activity is a key part in reducing GHG emissions; however, given the highly regional nature of transportation methods (i.e. high-speed trains are only an option in Europe, whereas many major US cities have limited public transportation), custom solutions are needed to make a significant impact. ML can be of particular help in estimating and predicting vehicle flow to minimize it, for example by helping to optimize the design of new roads and hubs~\citep{sommer2017} and monitoring traffic~\citep{kaack2019}, as well as estimating carbon emissions in real-time~\citep{nocera2018}. ML can also be used for designing more energy-efficient batteries~\citep{hoffmann2019} which will become an increasingly important concern as more people switch to electric vehicles. In both cases of energy and transportation, ML can be used to make systems more efficient and to improve predictions of complex phenomena based on large amounts of data; nonetheless, it remains only one part of the solution, and as tempting as it is to halt research projects once a theoretically plausible solution has been found (and a research paper published), what is key here is working with domain experts to bring projects towards deployment, where concrete impact can be made. Transversal connections between disciplines are therefore key, and must be established and fostered for projects to flourish.

\paragraph{Individuals and Societies}\mbox{} \\

While changes in our climate can be abstract, quantified in degrees of warming or tons of CO2, climate change will also have very concrete impacts on society, for instance by decreasing crop yield, increasing the frequency of extreme weather events such as hurricanes and storms, and impacting biodiversity. There are a myriad of ways in which ML can help face these, whether it be by analyzing real-time images and recordings of ecosystems to detect species~\citep{duhart2018} and deforestation~\citep{mcdowell2015}, improving disaster preparation and response by generating real-time maps from satellite imagery~\citep{voigt2007} and even setting an optimal price on carbon to accelerate the transition to a low-carbon energy economy~\citep{wei2018}. Finally, while we are far from being able to predict the exact impact that increasing the carbon tax will have on the different levels of society and industry (i.e. federal and regional governments, local and international companies, and individuals), this is a worthwhile area of research and exploration, with potentially huge consequences in helping political leaders make more informed choices in addressing the climate crisis. It is therefore useful to continue gathering data and building trust between members of the political ecosystem and ML practitioners to learn from each other and to facilitate the deployment of technological solutions in setting government policies.

On an individual level, there are many reasons why individuals cannot, or will not, act on climate change, either common misconceptions regarding the fact that individuals cannot make meaningful impact on a global problem, or cognitive biases that increase an individual’s psychological distance to climate change. In the first case, ML-infused tools to estimate the carbon footprint of individuals and households~\citep{jones2011} and to model individual behavior with regards to sustainable lifestyle choices and technologies~\citep{carr2011} can be very useful if they are sufficiently accurate and deployed on a large scale. Finally, minimizing psychological distance to the future effects of climate change is a promising way to reduce cognitive bias -- in this regard, it is possible to use images generated using Generative Adversarial Networks (GANs) which represent the impacts of extreme events on locations that have personal value to the viewer~\citep{schmidt2019}. A crucial part of developing ML tools for individuals is, once again, working with multidisciplinary experts in psychology, scientific communication, and user design to ensure that the tools created reach the largest possible audience and maximize their positive impact. 

\section{Conclusion}

Technology in general, and ML more specifically, carries a great potential for change and disruption. While neither of these is guaranteed to make the world a better place, this potential can most definitely be used to have a positive impact on the world. In the present article, we have illustrated some inspiring projects that aim to make the world a better place and by using the powerful techniques and approaches that ML has brought forward. We believe that as ML researchers and practitioners, we have the responsibility to leverage our (super)powers to contribute to these efforts. This can be done by connecting with established actors from industry and policy or experts from other relevant disciplines, by learning from their past experiences, and by working together to propose innovative solutions to major problems, deployed in places where they will have a positive impact. 

We live in a world with many global and local challenges and issues that are in constant evolution, and it is easy to be overwhelmed by this flux of information and focus on a small sandbox in which we feel safe and in control, in order to develop and study the aspects of ML that interest us most. But it is naive to believe that our sandbox is an isolated isle that is not connected to the rest of the world -- since even in the case of theoretical work, communication and cross-pollination are unavoidable -- and each of us is also a citizen who is concerned collective debates, while many of us could worry about the world in which our descendants will live. We believe that there are thought processes that should take place in the head of every ML practitioner regarding the nature of the work they are doing and the potential pitfalls and impacts of this work in the world around them, some of which we have listed in the first part of the current paper. And while we do not claim to have all the answers to all of these tough questions, we hope that we can start a conversation that will accompany ML research and practice throughout its infancy towards its tumultuous teenage years in the coming decades, and eventually towards mature adulthood beyond that. 

\bibliographystyle{plainnat}
\bibliography{morality}

\begin{thebibliography}{48}
\providecommand{\natexlab}[1]{#1}
\providecommand{\url}[1]{\texttt{#1}}
\expandafter\ifx\csname urlstyle\endcsname\relax
  \providecommand{\doi}[1]{doi: #1}\else
  \providecommand{\doi}{doi: \begingroup \urlstyle{rm}\Url}\fi

\bibitem[Abbott(2006)]{abbott2006}
Robert~G Abbott.
\newblock Automated expert modeling for automated student evaluation.
\newblock In \emph{International Conference on Intelligent Tutoring Systems},
  pages 1--10. Springer, 2006.

\bibitem[Allen et~al.(2019)Allen, Antwi-Agyei, Aragon-Durand, Babiker,
  Bertoldi, Bind, Brown, Buckeridge, Camilloni, Cartwright, et~al.]{allen2019}
M~Allen, P~Antwi-Agyei, F~Aragon-Durand, M~Babiker, P~Bertoldi, M~Bind,
  S~Brown, M~Buckeridge, I~Camilloni, A~Cartwright, et~al.
\newblock Technical summary: Global warming of 1.5c. an ipcc special report on
  the impacts of global warming of 1.5c above pre-industrial levels and related
  global greenhouse gas emission pathways, in the context of strengthening the
  global response to the threat of climate change, sustainable development, and
  efforts to eradicate poverty, 2019.

\bibitem[Angwin et~al.(2016)Angwin, Larson, Mattu, and Kirchner]{angwin2016}
Julia Angwin, Jeff Larson, Surya Mattu, and Lauren Kirchner.
\newblock Machine bias, propublica.
\newblock
  \url{https://www.propublica.org/article/machine-bias-risk-assessments-in-criminal-sentencing},
  2016.
\newblock Accessed: 2019-11-25.

\bibitem[Arcadu et~al.(2019)Arcadu, Benmansour, Maunz, Willis, Haskova, and
  Prunotto]{arcadu2019}
Filippo Arcadu, Fethallah Benmansour, Andreas Maunz, Jeff Willis, Zdenka
  Haskova, and Marco Prunotto.
\newblock Deep learning algorithm predicts diabetic retinopathy progression in
  individual patients.
\newblock \emph{NPJ digital medicine}, 2\penalty0 (1):\penalty0 1--9, 2019.

\bibitem[Assembly(1948)]{un1948}
UN~General Assembly.
\newblock Universal declaration of human rights.
\newblock \emph{UN General Assembly}, 302\penalty0 (2), 1948.

\bibitem[Bolukbasi et~al.(2016)Bolukbasi, Chang, Zou, Saligrama, and
  Kalai]{bolukbasi2016}
Tolga Bolukbasi, Kai-Wei Chang, James~Y Zou, Venkatesh Saligrama, and Adam~T
  Kalai.
\newblock Man is to computer programmer as woman is to homemaker? debiasing
  word embeddings.
\newblock In \emph{Advances in neural information processing systems}, pages
  4349--4357, 2016.

\bibitem[Buolamwini and Gebru(2018)]{buolamwini2018}
Joy Buolamwini and Timnit Gebru.
\newblock Gender shades: Intersectional accuracy disparities in commercial
  gender classification.
\newblock In \emph{Conference on fairness, accountability and transparency},
  pages 77--91, 2018.

\bibitem[Carr-Cornish et~al.(2011)Carr-Cornish, Ashworth, Gardner, and
  Fraser]{carr2011}
Simone Carr-Cornish, Peta Ashworth, John Gardner, and Stephen~J Fraser.
\newblock Exploring the orientations which characterise the likely public
  acceptance of low emission energy technologies.
\newblock \emph{Climatic change}, 107\penalty0 (3-4):\penalty0 549--565, 2011.

\bibitem[Chaplot et~al.(2015)Chaplot, Rhim, and Kim]{chaplot2015}
Devendra~Singh Chaplot, Eunhee Rhim, and Jihie Kim.
\newblock Predicting student attrition in moocs using sentiment analysis and
  neural networks.
\newblock In \emph{AIED Workshops}, volume~53, pages 54--57, 2015.

\bibitem[Conati et~al.(2018)Conati, Porayska-Pomsta, and Mavrikis]{conati2018}
Cristina Conati, Kaska Porayska-Pomsta, and Manolis Mavrikis.
\newblock Ai in education needs interpretable machine learning: Lessons from
  open learner modelling.
\newblock \emph{arXiv preprint arXiv:1807.00154}, 2018.

\bibitem[Dastin(2018)]{dastin2018}
Jeffrey Dastin.
\newblock Amazon scraps secret ai recruiting tool that showed bias against
  women, reuters business news.
\newblock
  \url{https://www.reuters.com/article/us-amazon-com-jobs-automation-insight/amazon-scraps-secret-ai-recruiting-tool-that-showed-bias-against-women-idUSKCN1MK08G},
  2018.
\newblock Accessed: 2019-11-25.

\bibitem[Deng et~al.(2009)Deng, Dong, Socher, Li, Li, and Fei-Fei]{deng2009}
Jia Deng, Wei Dong, Richard Socher, Li-Jia Li, Kai Li, and Li~Fei-Fei.
\newblock Imagenet: A large-scale hierarchical image database.
\newblock In \emph{2009 IEEE conference on computer vision and pattern
  recognition}, pages 248--255. Ieee, 2009.

\bibitem[Duhart et~al.(2018)Duhart, Dublon, Mayton, and Paradiso]{duhart2018}
Clement Duhart, Gershon Dublon, Brian Mayton, and Joseph Paradiso.
\newblock Deep learning locally trained wildlife sensing in real acoustic
  wetland environment.
\newblock In \emph{International Symposium on Signal Processing and Intelligent
  Recognition Systems}, pages 3--14. Springer, 2018.

\bibitem[Dutt et~al.(2017)Dutt, Ismail, and Herawan]{dutt2017}
Ashish Dutt, Maizatul~Akmar Ismail, and Tutut Herawan.
\newblock A systematic review on educational data mining.
\newblock \emph{IEEE Access}, 5:\penalty0 15991--16005, 2017.

\bibitem[Ghassemi et~al.(2018)Ghassemi, Naumann, Schulam, Beam, and
  Ranganath]{ghassemi2018}
Marzyeh Ghassemi, Tristan Naumann, Peter Schulam, Andrew~L Beam, and Rajesh
  Ranganath.
\newblock Opportunities in machine learning for healthcare.
\newblock \emph{arXiv preprint arXiv:1806.00388}, 2018.

\bibitem[Golden(2017)]{golden2017}
Jeffrey~Alan Golden.
\newblock Deep learning algorithms for detection of lymph node metastases from
  breast cancer: helping artificial intelligence be seen.
\newblock \emph{Jama}, 318\penalty0 (22):\penalty0 2184--2186, 2017.

\bibitem[Goodfellow et~al.(2014)Goodfellow, Pouget-Abadie, Mirza, Xu,
  Warde-Farley, Ozair, Courville, and Bengio]{goodfellow2014}
Ian Goodfellow, Jean Pouget-Abadie, Mehdi Mirza, Bing Xu, David Warde-Farley,
  Sherjil Ozair, Aaron Courville, and Yoshua Bengio.
\newblock Generative adversarial nets.
\newblock In \emph{Advances in neural information processing systems}, pages
  2672--2680, 2014.

\bibitem[Gotterbarn et~al.(2018)Gotterbarn, Bruckman, Flick, Miller, and
  Wolf]{gotterbarn2018}
Don~W Gotterbarn, Amy Bruckman, Catherine Flick, Keith Miller, and Marty~J
  Wolf.
\newblock Acm code of ethics: a guide for positive action, 2018.

\bibitem[Head et~al.(2008)]{head2008}
Brian~W Head et~al.
\newblock Wicked problems in public policy.
\newblock \emph{Public policy}, 3\penalty0 (2):\penalty0 101, 2008.

\bibitem[Hitti et~al.(2019)Hitti, Jang, Moreno, and Pelletier]{hitti2019}
Yasmeen Hitti, Eunbee Jang, Ines Moreno, and Carolyne Pelletier.
\newblock Proposed taxonomy for gender bias in text; a filtering methodology
  for the gender generalization subtype.
\newblock In \emph{Proceedings of the First Workshop on Gender Bias in Natural
  Language Processing}, pages 8--17, 2019.

\bibitem[Hoffmann et~al.(2019)Hoffmann, Maestrati, Sawada, Tang, Sellier, and
  Bengio]{hoffmann2019}
Jordan Hoffmann, Louis Maestrati, Yoshihide Sawada, Jian Tang, Jean~Michel
  Sellier, and Yoshua Bengio.
\newblock Data-driven approach to encoding and decoding 3-d crystal structures.
\newblock \emph{arXiv preprint arXiv:1909.00949}, 2019.

\bibitem[IEEE(2017)]{ieee2017}
IEEE.
\newblock Ieee standard review — ethically aligned design: A vision for
  prioritizing human wellbeing with artificial intelligence and autonomous
  systems.
\newblock In \emph{2017 IEEE Canada International Humanitarian Technology
  Conference (IHTC)}, pages 197--201. IEEE, 2017.

\bibitem[Jobin et~al.(2019)Jobin, Ienca, and Vayena]{jobin2019}
Anna Jobin, Marcello Ienca, and Effy Vayena.
\newblock Artificial intelligence: the global landscape of ethics guidelines.
\newblock \emph{arXiv preprint arXiv:1906.11668}, 2019.

\bibitem[Jones and Kammen(2011)]{jones2011}
Christopher~M Jones and Daniel~M Kammen.
\newblock Quantifying carbon footprint reduction opportunities for us
  households and communities.
\newblock \emph{Environmental science \& technology}, 45\penalty0 (9):\penalty0
  4088--4095, 2011.

\bibitem[Kaack et~al.(2019)Kaack, Chen, and Morgan]{kaack2019}
Lynn~H Kaack, George~H Chen, and M~Granger Morgan.
\newblock Truck traffic monitoring with satellite images.
\newblock In \emph{Proceedings of the Conference on Computing \& Sustainable
  Societies}, pages 155--164. ACM, 2019.

\bibitem[LeCun et~al.(1998)LeCun, Bottou, Bengio, Haffner, et~al.]{lecun1998}
Yann LeCun, L{\'e}on Bottou, Yoshua Bengio, Patrick Haffner, et~al.
\newblock Gradient-based learning applied to document recognition.
\newblock \emph{Proceedings of the IEEE}, 86\penalty0 (11):\penalty0
  2278--2324, 1998.

\bibitem[Levin et~al.(2012)Levin, Cashore, Bernstein, and Auld]{levin2012}
Kelly Levin, Benjamin Cashore, Steven Bernstein, and Graeme Auld.
\newblock Overcoming the tragedy of super wicked problems: constraining our
  future selves to ameliorate global climate change.
\newblock \emph{Policy sciences}, 45\penalty0 (2):\penalty0 123--152, 2012.

\bibitem[McDowell et~al.(2015)McDowell, Coops, Beck, Chambers, Gangodagamage,
  Hicke, Huang, Kennedy, Krofcheck, Litvak, et~al.]{mcdowell2015}
Nate~G McDowell, Nicholas~C Coops, Pieter~SA Beck, Jeffrey~Q Chambers, Chandana
  Gangodagamage, Jeffrey~A Hicke, Cho-ying Huang, Robert Kennedy, Dan~J
  Krofcheck, Marcy Litvak, et~al.
\newblock Global satellite monitoring of climate-induced vegetation
  disturbances.
\newblock \emph{Trends in plant science}, 20\penalty0 (2):\penalty0 114--123,
  2015.

\bibitem[Nkambou et~al.(2010)Nkambou, Mizoguchi, and Bourdeau]{nkambou2010}
Roger Nkambou, Riichiro Mizoguchi, and Jacqueline Bourdeau.
\newblock \emph{Advances in intelligent tutoring systems}, volume 308.
\newblock Springer Science \& Business Media, 2010.

\bibitem[Nocera et~al.(2018)Nocera, Ruiz-Alarc{\'o}n-Quintero, and
  Cavallaro]{nocera2018}
Silvio Nocera, Cayetano Ruiz-Alarc{\'o}n-Quintero, and Federico Cavallaro.
\newblock Assessing carbon emissions from road transport through traffic flow
  estimators.
\newblock \emph{Transportation Research Part C: Emerging Technologies},
  95:\penalty0 125--148, 2018.

\bibitem[Nye(2015)]{nye2015}
Benjamin~D Nye.
\newblock Intelligent tutoring systems by and for the developing world: A
  review of trends and approaches for educational technology in a global
  context.
\newblock \emph{International Journal of Artificial Intelligence in Education},
  25\penalty0 (2):\penalty0 177--203, 2015.

\bibitem[Perisic(2018)]{wef2018}
I~Perisic.
\newblock How artificial intelligence is shaking up the job market.
\newblock
  \url{https://www.weforum.org/agenda/2018/09/artificial-intelligence-shaking-up-job-market/},
  2018.
\newblock Accessed: 2019-11-25.

\bibitem[Pivovarov et~al.(2015)Pivovarov, Perotte, Grave, Angiolillo, Wiggins,
  and Elhadad]{pivovarov2015}
Rimma Pivovarov, Adler~J Perotte, Edouard Grave, John Angiolillo, Chris~H
  Wiggins, and No{\'e}mie Elhadad.
\newblock Learning probabilistic phenotypes from heterogeneous ehr data.
\newblock \emph{Journal of biomedical informatics}, 58:\penalty0 156--165,
  2015.

\bibitem[Ribeiro et~al.(2016)Ribeiro, Singh, and Guestrin]{ribeiro2016}
Marco~Tulio Ribeiro, Sameer Singh, and Carlos Guestrin.
\newblock Why should i trust you?: Explaining the predictions of any
  classifier.
\newblock In \emph{Proceedings of the 22nd ACM SIGKDD international conference
  on knowledge discovery and data mining}, pages 1135--1144. ACM, 2016.

\bibitem[Rolnick et~al.(2019)Rolnick, Donti, Kaack, Kochanski, Lacoste,
  Sankaran, Ross, Milojevic-Dupont, Jaques, Waldman-Brown, Luccioni,
  et~al.]{rolnick2019}
David Rolnick, Priya~L Donti, Lynn~H Kaack, Kelly Kochanski, Alexandre Lacoste,
  Kris Sankaran, Andrew~Slavin Ross, Nikola Milojevic-Dupont, Natasha Jaques,
  Anna Waldman-Brown, Alexandra Luccioni, et~al.
\newblock Tackling climate change with machine learning.
\newblock \emph{arXiv preprint arXiv:1906.05433}, 2019.

\bibitem[RSS(2019)]{rss2019}
RSS.
\newblock A guide for ethical data science.
\newblock
  \url{https://www.actuaries.org.uk/system/files/field/document/An\%20Ethical\%20Charter\%20for\%20Date\%20Science\%20WEB\%20FINAL.PDF.},
  2019.
\newblock Accessed: 2019-11-25.

\bibitem[Schmidt et~al.(2019)Schmidt, Luccioni, Mukkavilli, Balasooriya,
  Sankaran, Chayes, and Bengio]{schmidt2019}
Victor Schmidt, Alexandra Luccioni, S~Karthik Mukkavilli, Narmada Balasooriya,
  Kris Sankaran, Jennifer Chayes, and Yoshua Bengio.
\newblock Visualizing the consequences of climate change using cycle-consistent
  adversarial networks.
\newblock \emph{arXiv preprint arXiv:1905.03709}, 2019.

\bibitem[Sommer et~al.(2017)Sommer, Schuchert, and Beyerer]{sommer2017}
Lars~Wilko Sommer, Tobias Schuchert, and J{\"u}rgen Beyerer.
\newblock Fast deep vehicle detection in aerial images.
\newblock In \emph{2017 IEEE Winter Conference on Applications of Computer
  Vision (WACV)}, pages 311--319. IEEE, 2017.

\bibitem[Suppes and Morningstar(1969)]{suppes1969}
Patrick Suppes and Mona Morningstar.
\newblock Computer-assisted instruction.
\newblock \emph{Science}, 166\penalty0 (3903):\penalty0 343--350, 1969.

\bibitem[Szczepański(2019)]{economicai2019}
M~Szczepański.
\newblock Economic impacts of artificial intelligence (ai), european
  parliamentary research service, pe 637.967.
\newblock
  \url{http://europarl.europa.eu/RegData/etudes/BRIE/2019/637967/EPRS_BRI(2019)637967_EN.pdf},
  2019.
\newblock Accessed: 2019-11-25.

\bibitem[UNHCR(2017)]{sdg2017}
UNHCR.
\newblock The sustainable development goals and addressing statelessness.
\newblock \url{https://www.refworld.org/docid/58b6e3364.html}, 2017.
\newblock Accessed: 2019-11-25.

\bibitem[V{\'a}zquez-Canteli and Nagy(2019)]{vazquez2019}
Jos{\'e}~R V{\'a}zquez-Canteli and Zolt{\'a}n Nagy.
\newblock Reinforcement learning for demand response: A review of algorithms
  and modeling techniques.
\newblock \emph{Applied energy}, 235:\penalty0 1072--1089, 2019.

\bibitem[Voigt et~al.(2007)Voigt, Kemper, Riedlinger, Kiefl, Scholte, and
  Mehl]{voigt2007}
Stefan Voigt, Thomas Kemper, Torsten Riedlinger, Ralph Kiefl, Klaas Scholte,
  and Harald Mehl.
\newblock Satellite image analysis for disaster and crisis-management support.
\newblock \emph{IEEE transactions on geoscience and remote sensing},
  45\penalty0 (6):\penalty0 1520--1528, 2007.

\bibitem[Voyant et~al.(2017)Voyant, Notton, Kalogirou, Nivet, Paoli, Motte, and
  Fouilloy]{voyant2017}
Cyril Voyant, Gilles Notton, Soteris Kalogirou, Marie-Laure Nivet, Christophe
  Paoli, Fabrice Motte, and Alexis Fouilloy.
\newblock Machine learning methods for solar radiation forecasting: A review.
\newblock \emph{Renewable Energy}, 105:\penalty0 569--582, 2017.

\bibitem[Wei et~al.(2018)Wei, Chongchong, and Cuiping]{wei2018}
Sun Wei, Zhang Chongchong, and Sun Cuiping.
\newblock Carbon pricing prediction based on wavelet transform and k-elm
  optimized by bat optimization algorithm in china ets: the case of shanghai
  and hubei carbon markets.
\newblock \emph{Carbon Management}, 9\penalty0 (6):\penalty0 605--617, 2018.

\bibitem[Wexler et~al.(2019)Wexler, Pushkarna, Bolukbasi, Wattenberg,
  Vi{\'e}gas, and Wilson]{wexler2019}
James Wexler, Mahima Pushkarna, Tolga Bolukbasi, Martin Wattenberg, Fernanda
  Vi{\'e}gas, and Jimbo Wilson.
\newblock The what-if tool: Interactive probing of machine learning models.
\newblock \emph{IEEE transactions on visualization and computer graphics},
  26\penalty0 (1):\penalty0 56--65, 2019.

\bibitem[Wiens et~al.(2019)Wiens, Saria, Sendak, Ghassemi, Liu, Doshi-Velez,
  Jung, Heller, Kale, Saeed, et~al.]{wiens2019}
Jenna Wiens, Suchi Saria, Mark Sendak, Marzyeh Ghassemi, Vincent~X Liu, Finale
  Doshi-Velez, Kenneth Jung, Katherine Heller, David Kale, Mohammed Saeed,
  et~al.
\newblock Do no harm: a roadmap for responsible machine learning for health
  care.
\newblock \emph{Nature medicine}, 25\penalty0 (9):\penalty0 1337--1340, 2019.

\bibitem[Zemel et~al.(2013)Zemel, Wu, Swersky, Pitassi, and Dwork]{zemel2013}
Rich Zemel, Yu~Wu, Kevin Swersky, Toni Pitassi, and Cynthia Dwork.
\newblock Learning fair representations.
\newblock In \emph{International Conference on Machine Learning}, pages
  325--333, 2013.

\end{thebibliography}

\end{document}